\documentstyle[aps,multicol,prb,epsfig]{revtex} 

\renewcommand{\caption}[1]{\refstepcounter{figure}\protect\noindent%
  \protect\parbox{8.6cm}{\small FIG. \thefigure. #1}}

\begin{document}
 
\title{A field-oriented chain of dipolar particles in elongational
  flow}
  
\author{M.-Carmen Miguel$^1$ and J.M. Rub\'{\i}$^2$}

\address{$^1$ Department of Physics\\
  Massachusetts Institute of Technology, Cambridge, Massachusetts
  02139\\
  $^2$ Departament de F\'{\i}sica Fonamental, Facultat de F\'{\i}sica
  \\ 
  Universitat de Barcelona\\ Diagonal 647, 08028 Barcelona, Spain\\
  \date{\today}}

\maketitle

\draft
   
\begin{abstract} 
 
  We study the behavior of an isolated field-oriented chain of dipolar
  particles in elongational fluid flow. Our main goal is to emphasize
  the effect of dipolar interactions on the chain's contribution to
  the pressure tensor and to the viscosities of a dilute suspension of
  these linear aggregates.  In our model, despite the overall rigid
  appearance of the chain at rest, the constituent beads may move
  slightly relative to one another, conferring a certain degree of
  flexibility to the chain.  This flexibility is quantified in terms
  of a dimensionless parameter, $\lambda^{-1}$, comparing thermal and
  dipolar energies. We perform an expansion in $\lambda^{-1}$, and
  obtain the first correction to the {\em rigid chain} contribution to
  the Kramers' pressure tensor for different flow geometries.  The
  interplay of the elongational flow field and the field-induced chain
  orientation gives rise to a rich variety of scenarios. We compute
  the elongational, shear, and rotational viscosities in some
  representative situations.

\end{abstract}


\begin{multicols}{2}
  
\section{Introduction}  

\label{intro}

Colloidal suspensions play an important role in many natural phenomena
as well as in various industrial processes. The stability of these
suspensions against aggregation of their constituent particles is of
essential importance for their behavior \cite{gast89}. As long as
interparticle interactions can be neglected, one does not observe
aggregation and the characterization of the system reduces to
understanding how the physical properties of the fluid,---e.g. its
viscosity---are modified due to the presence of the particles
\cite{VdV89,shliomis72,miguel95,bacri95}. On the contrary, the
phenomenology of the suspension may change dramatically if
interactions between particles become relevant.

In the late 1930's, Winslow \cite{winslow49} first noted the curious
behavior of a suspension of dielectric particles in oil when subjected
to an electric field. He reported the formation of linear chains of
particles aligned with the electric field, and how the effective
viscosity of the suspension could change by orders of magnitude by
simply modifying the applied field. An analogous field-induced
behavior is exhibited by magnetorheological fluids, e.g., by a
suspension of magnetizable superparamagnetic particles in a
nonmagnetic fluid \cite{bossis90}, or by a suspension of
nonmagnetizable spheres in a ferrofluid (magnetic holes)
\cite{skjeltorp85,davies86,ferrofluid}.  Since then, several
experimental and theoretical efforts, as well as computer simulations,
have been devoted to study the different aspects of the complex
behavior of colloidal suspensions of dipolar particles
\cite{miyazima87,helgesen88,bossis89,fraden89,fermigier92,klingenberg93,lemaire94,gast95,miguel98}.

The change in the rheological properties of dipolar suspensions upon
the action of an external field, is in part due to the aggregation of
the colloidal particles, which form clusters of macroscopic size.
These are usually linear chains---quasi rigid rods---oriented along
the direction of the applied field, although for high enough
concentrations of dipolar particles more complex structures may arise
\cite{deGennes70,wang94,jund95}. The overall spatial arrangement of
the aggregates is very effective in hindering the fluid flow,
conferring the suspension a solid-like texture.

Our purpose in this paper is to ascertain the role played by strong,
but finite, dipolar interactions in the linear rheology of a
suspension containing either an isolated field-induced aggregate or,
equivalently, a dilute concentration of such clusters. After
introducing the dimensionless parameter, $\lambda$, comparing dipolar
and thermal energies, we obtain the first correction in powers of
$\lambda^{-1}$ to the {\em rigid rod} ($\lambda^{-1}\rightarrow 0$)
contribution to the Kramers' pressure tensor for different flow
geometries. In particular, we analyze the influence that those
interactions exert on the value of some of the transport coefficients
of the system.  We have organized the paper in the following way:
Section \ref{description} contains a description of the system and the
conditions under study. The {\em end-to-end} vector of the chain is
introduced and its mean value, as well as the mean square {\em
  end-to-end} distance, are obtained.  In Section \ref{pressure} we
analyze the contribution of a chain, under the action of an
elongational flow, to the pressure tensor of the suspension, and to
some of the viscosities characterizing the system for different
elongational flow fields of interest. The conclusions are summed up in
the last section.
 
\section{Description of the system}

\label{description}

The structure of a colloidal particle aggregate may be either fixed or
deformable, depending upon the nature of the aggregation process and
the type of interparticle bonds established. In
Ref.~\onlinecite{fraden89}, the issue of the growth of similar
electric field-induced chains was experimentally addressed. There, the
authors reported that, in the presence of an intense field, the
colloidal dipoles organize themselves into linear chains. These
clusters exhibit a certain degree of flexibility, limited of course by
the strength of the dipolar interaction among the beads. Within one
chain, they observed fluctuations in the separation between sphere
surfaces ($5-10\%$ of a diameter), as well as in the angle between the
relative position vector of consecutive spheres and the applied field.
Hence, we will assume that the beads forming the chain are in close
contact, but still able to slightly move in relation to one another
due to thermal fluctuations and to the action of an elongational flow
field.

The dimensionless parameters describing the conditions of the system
under study are essentially: $\lambda= m^2/(d^3 k_B T)$, comparing
dipolar and thermal energies, and the Langevin parameter $\mu\equiv
mH/k_BT$, where $m$ and $d$ are the dipolar moment and the diameter of
each sphere, $H$ is the external (magnetic or electric) field
strength, $k_B$ is the Boltzmann constant, and $T$ the absolute
temperature \cite{ferrofluid2}.  In our analysis $\lambda$ is assumed
to be large and $\mu \rightarrow \infty$, as in the experimental
realizations in which the irreversible formation of these sort of
quasi rigid linear aggregates is observed.

Despite the long-range contributions to the energy of a chain
containing a fixed number of particles, $N$, the amplitude of the
vibrations and oscillations of the spheres are almost entirely taken
into account by just considering interactions between
nearest-neighbors \cite{deGennes70,jund95}. Long range effects are
subleading corrections to the short range attractive contributions and
can be ignored.  As a matter of fact, the most important contributions
correspond to relative distances among the particles of the order of
one diameter, and relative orientations with an azimuthal angle,
defined with respect to the external field direction, close to
$\theta=0$.

Thus we will assume that the potential energy can be expressed as a
sum of nearest neighbors terms

\begin{equation}\label{ch5}
\frac{V_{mag}}{k_BT}=\sum_{i=1}^{N-1} \Phi_{i,i+1}.
\end{equation}

\noindent Here,  

\begin{equation}\label{ch5b}
\Phi_{i,j}=\frac{\bbox{m}_i\cdot \bbox{m}_j - 3(\bbox{m}_i\cdot
 \hat{\bbox{r}}_{ij})(\bbox{m}_j\cdot \hat{\bbox{r}}_{ij})}{r_{ij}^3}
\end{equation}
 
\noindent is the interaction energy of the dipoles $i$ and $j$, with  
$\bbox{r}_{ij}=r_{ij}\hat{\bbox{r}}_{ij}$ the vector giving their relative
position.  Its modulus $r_{ij}$ is the distance between the sphere
centers. Since the most important contributions come from distances
$r_{ij}\sim d$ and relative orientations $\theta_{ij}\sim 0$, each
term in the sum of Eq.~(\ref{ch5}) can be approximately written as
follows \cite{deGennes70}:

\begin{equation}\label{ch6}
\Phi_{i,i+1} \sim - \lambda (2-3 \theta_i^2- 6 \xi_i).
\end{equation}

\begin{figure}[t]
\centerline{\epsfig{file=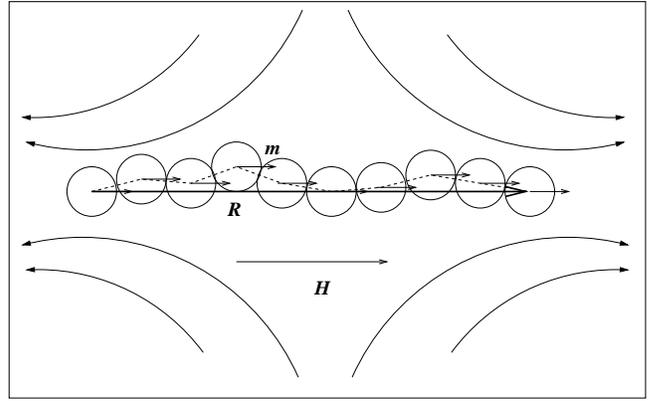, width=8.5cm}} 
\vspace*{0.5cm}
\caption{Chain of magnetic particles in an elongational 
flow. All the magnetic moments $\bbox{m}$ point towards the direction 
of the external field applied. The vector $\bbox{R}$ is the 
{\em end-to-end} vector giving the relative position of the 
centers of the spheres on the extremes.}
\label{fig1}
\end{figure}

\noindent In the last equation, we have introduced the relative
position vector between two consecutive spheres in the chain
$\bbox{q}_i=\bbox{R}_{i+1}-\bbox{R}_{i}$ through its components in
spherical coordinates $\bbox{q}_i\equiv(q_i,\theta_i,\varphi_i)$. The
position vector $\bbox{R}_i$ determines the center of the $i$-th
sphere. Moreover, we have performed an expansion around the minimum of
the energy in the variables $\theta_i$ and $\xi_i$, which, in turn, is
defined by $q_i = d(1+\xi_i)$, and such that $0\leq \xi_i \ll 1$.

Multipole and multibody contributions are not considered because they
are expected to be very small. Likewise, we disregard London-Van der
Waals forces, and any possible steric interactions due to the
surfactant monolayer with which the colloidal particles may be usually
coated.

\subsection*{Mean size of the chain}

To characterize the mean size of a chain, containing $N$ dipoles,
under the action of an elongational flow and thermal fluctuations, let
us consider the {\em end-to-end} vector $\bbox{R}$ giving the relative
position of the centers of the spheres located on the tips of the
chain (see Fig.~\ref{fig1})

\begin{equation}\label{ch7b}
\bbox{R}=\bbox{R}_N - \bbox{R}_1= \sum_{k=1}^{N-1} \bbox{q}_k.
\end{equation}

\noindent In order to compute the average of this vector, or
equivalently of the vector $\bbox{q}_k$, we need the probability
distribution $\psi(\{\bbox{R}_i\})$ of the spheres' position.
This probability is the solution of the diffusion equation describing
the conservation of system points in the configurational space
\cite{rotation},

\begin{eqnarray}\label{ch2}
& &\frac{\partial \psi}{\partial t}= \sum_i \left\{ 
\frac{\partial}{\partial \bbox{R}_i} \cdot \sum_j \bbox{\mu}_{ij}^{TT}
  \cdot \left( k_BT \frac{\partial \psi}{\partial \bbox{R}_j} +
  \frac{\partial  
V_{mag}}{\partial \bbox{R}_j} \psi \right) \right.\nonumber \\
& & \mbox{\ \ } \left. - \frac{\partial}{\partial
\bbox{R}_i} 
\cdot (\bbox{\beta}\cdot \bbox{R}_i) \psi \right\},
\end{eqnarray}

\noindent where $\bbox{\mu}_{ij}^{TT}$ is the relative
translational mobility tensor, $V_{mag}$ is the dipole-dipole magnetic
potential, and $\bbox{\beta}$ is the elongational rate
tensor ($\bbox{\beta}=\bbox{\beta}^T$)
corresponding to the stationary homogeneous external flow,
$\bbox{v}_o=\bbox{\beta}\cdot \bbox{r}$.  As an
approximation, we assume that the spheres move through the solvent
without disturbing the velocity field ({\it free draining}), so that

\begin{equation}\label{ch3}
\bbox{\mu}_{ij}^{TT}=\frac{1}{6\pi\eta a}\bbox{I} \ 
\delta_{ij},
\end{equation}

\noindent where, for the sake of simplicity, we consider that all the
spheres in the chain have the same radius $a=d/2$, and $\bbox{I}$ is the
unit tensor.

The next step is to write the diffusion equation in terms of the
relative position vectors, $\bbox{q}_i$. To this purpose we introduce
the transformation \cite{bird76}

\begin{equation}\label{ch8}
\bbox{q}_i=\sum_k B_{ik} \bbox{R}_k, \qquad \mbox{with} \qquad
B_{ik}= \delta_{i+1,k}-\delta_{i,k}, 
\end{equation}

\noindent from which one infers the relation $\partial/\partial
\bbox{R}_i=\sum_k B_{ki}\ \partial/\partial \bbox{q}_k$; and the
matrix

\begin{equation}\label{ch10}
 A_{ij}=\sum_k B_{ik} B_{jk}=\left\{ \begin{array}{lll}
2 & \mbox{if}\  i=j \nonumber\\
-1 & \mbox{if}\  i=j\pm 1 \nonumber\\
0 & \mbox{otherwise}\nonumber
\end{array}\right..
\end{equation}

\noindent Taking into account Eqs.~(\ref{ch3})-(\ref{ch10}), the
diffusion equation can be alternatively written as

\begin{eqnarray}\label{ch11}
& &\frac{\partial \psi}{\partial t}= \frac{k_BT}{6\pi\eta a} \sum_j 
\frac{\partial}{\partial \bbox{q}_j} \cdot \sum_k A_{jk}
 \left(\frac{\partial \psi}{\partial \bbox{q}_k} + 
\frac{\partial V_{mag}/k_BT}{\partial \bbox{q}_k} \psi \right) 
\nonumber \\
& & \mbox{\ \ } - 
\sum_j\frac{\partial}{\partial \bbox{q}_j} \cdot (\bbox{\beta} \cdot
\bbox{q}_j)   \psi. 
\end{eqnarray}

The stationary solution of this equation with a homogeneous
potential flow field is 

\begin{eqnarray}\label{ch12}
& &\psi_{st}(\bbox{q}_1,\ldots,\bbox{q}_{N-1}) \sim {\rm exp} \left\{ 
\frac{-V_{mag}}{k_BT} \right. \nonumber \\
& & \mbox{\ \ \ \ } \left. +\frac{3\pi\eta a}{k_BT}(\bbox{\beta}:\sum_i  
\sum_j C_{ij} \ \bbox{q}_i \ \bbox{q}_j) \right\},
\end{eqnarray}

\noindent where the symbol $:$ indicates a double contraction of
indexes, and $C_{ij}$ is the Kramers matrix defined in
Ref.~\onlinecite{bird76}. Up to linear order in the elongational rate
$\bbox{\beta}$---linear dynamics---, we expand the exponential factor
in Eq.~(\ref{ch12}).  The stationary average of $\bbox{q}_k$ is then
given by

\begin{eqnarray}\label{ch18b}
& & \langle \bbox{q}_k \rangle= \int 
(\prod_{l=1}^{N-1} 
d \bbox{q}_l)\ \bbox{q}_k\  \psi_{st} \simeq d \hat{\bbox{e}}_z \nonumber \\
& & \mbox{\ \ \ \ } + \frac{d^3}{6D} \lambda^{-1} 
(\bbox{\beta}\cdot\hat{\bbox{e}}_z - \beta_{zz} \hat{\bbox{e}}_z) \sum_i
C_{ik}. 
\end{eqnarray}

\noindent where we have introduced the translational diffusion
coefficient for a single particle $D=k_BT/(6\pi\eta a)$. This result
is obtained after decomposing the integral into different parts, and
taking into account the following relations \cite{bird76}:

\begin{equation}\label{ch17b}
\sum_{i=1}^{N-1} \sum_{j=1}^{N-1} C_{ij}= \frac{N(N^2-1)}{12}, \qquad
\sum_{i=1}^{N-1} C_{ii} = \frac{N^2-1}{6}. 
\end{equation} 

\noindent In the Appendix we indicate in more detail some of the
calculations involved in the computation of a similar average which
appears in the following section.

Therefore, up to first order in $\lambda^{-1}$ and in the elongational
flow rate, the average {\em end-to-end} vector $\bbox{R}$ defined in
Eq.~(\ref{ch7b}) turns out to be

\begin{equation}\label{ch18d}
\langle \bbox{R} \rangle \simeq (N-1) d \hat{\bbox{e}}_z + 
N(N^2-1)\frac{d^3}{72D}\lambda^{-1}
(\bbox{\beta}\cdot\hat{\bbox{e}}_z - \beta_{zz} \hat{\bbox{e}}_z),
\end{equation}

\noindent where, once again, use has been made of the relations
(\ref{ch17b}). At equilibrium ($\bbox{\beta}=0$), this average
obviously reduces to $\langle\bbox{R}\rangle_{eq}= (N-1) d
\hat{\bbox{e}}_z$---the length of a rigid chain.

Following the same procedure we can compute the mean square {\em
  end-to-end} distance

\begin{equation}\label{ch18e}
\langle R^2 \rangle=\langle \bbox{R}\cdot\bbox{R} \rangle=
\sum_{k=1}^{N-1} 
\sum_{m=1}^{N-1} \langle \bbox{q}_k\cdot\bbox{q}_m \rangle.
\end{equation}

\noindent Up to first order in $\beta$, and in the small parameter 
$\lambda^{-1}$ of our expansion, we do not find any contribution from
the elongational rate. Nevertheless, the equilibrium value is given by

\begin{equation}\label{ch18f}
\langle R^2 \rangle_{eq} \simeq (N-1)^2 d^2 + \frac{(N-1)}{3} \lambda^{-1}
d^2. 
\end{equation}

\noindent Thermal fluctuations effectively stretch a quasi rigid
field-oriented chain---using the root mean square end-to-end distance
as a measure of its length---by a global amount $[(N-1)/(3\lambda)]^{1/2}
d$.  On the other hand, thermal effects would be obviously negligible for a
rigid chain ($\lambda \gg N$) \cite{rigid}. The stretching factor per
bond is

\begin{equation}\label{ch18g}
\frac{[\langle R^2 \rangle_{eq}-\langle \bbox{R} \rangle_{eq}^2]^{1/2}}{N-1}=
[3\lambda(N-1)]^{-1/2}\ d.
\end{equation}

\noindent  Thus, for large but finite values of $\lambda N$,
i.e. $(\lambda N)^{-1}\neq 0$, the mean separation between consecutive
spheres increases by a factor proportional to $(\lambda N)^{-1/2}$.
This result (\ref{ch18g}) is in qualitative and quantitative agreement
with the observations reported in Ref.~\onlinecite{fraden89} for
similar electric field-induced chains.  In these experiments, 
$\lambda\sim (20-30)$ and the mean number of particles in a chain at
the late stages of the aggregation process is typically $N\sim
(5-15)$ which, according to our result (\ref{ch18g}), yield a mean
near-neighbors separation of about $(3-6)\%$ of $d$. This value is
comparable to their observations, in which the sphere surfaces appear 
to be separated by $(5-10)\%$ of a sphere diameter due to thermal 
fluctuations.
 
\section{Contribution of the chain to the pressure tensor}

\label{pressure}

Besides the contribution of the solvent to the total pressure tensor
of the suspension, there is another contribution coming from the
direct interaction of the particles constituting the chains. The
latter can be obtained from the rheological equation of state proposed
by Kramers \cite{bird76,doi89}

\begin{equation}\label{ch1}
\bbox{\Pi}^p= - \frac{1}{V} \sum_{k=1}^{N-1} \left\langle
  \bbox{q}_k \frac{\partial V_{mag}}{\partial \bbox{q}_k} \right\rangle
+ \frac{N-1}{V} k_BT \bbox{I},
\end{equation}

\noindent where $V$ is the volume of the system.

Once we know the stationary solution of the diffusion equation
(\ref{ch12}), the average appearing on the right hand side of equation
(\ref{ch1}) can be expressed as an integral of the form:

\begin{equation}\label{ch14}
\left\langle \bbox{q}_k 
\frac{\partial V_{mag}}{\partial \bbox{q}_k} \right\rangle= \int 
\prod_{l=1}^{N-1} d 
\bbox{q}_l \left(\bbox{q}_k 
\frac{\partial V_{mag}}{\partial \bbox{q}_k} 
\right)\psi_{st}(\bbox{q}_1,\ldots,\bbox{q}_{N-1}).
\end{equation}
 
\noindent As we are interested in finding the Newtonian
viscosity tensor, we again expand the exponential factor in
Eq.~(\ref{ch12}) up to first order in $\bbox{\beta}$. To
evaluate the remaining integral, it is convenient to decompose it into
different parts as indicated in the Appendix; thereby, in the
Newtonian domain, the average in Eq.~(\ref{ch14}) reads

\begin{eqnarray}\label{ch19}
& &\left\langle \bbox{q}_k 
\frac{\partial V_{mag}}{\partial \bbox{q}_k} \right\rangle \simeq k_B
T  \left\{\bbox{I} +\frac{d^2}{D} \sum_i C_{ik} \bbox{\beta}
\cdot \hat{\bbox{e}}_z\hat{\bbox{e}}_z \right. \nonumber \\
& & \mbox{\ \ \ } \left. +\frac{d^2}{6D}\lambda^{-1}
C_{kk} \bbox{\beta}\cdot(\bbox{I} 
-\hat{\bbox{e}}_z\hat{\bbox{e}}_z)\right\}.
\end{eqnarray}

\noindent Finally, after using Eqs.~(\ref{ch17b}), we obtain the
contribution of one chain to the pressure tensor (\ref{ch1})

\begin{equation}\label{ch20}
\bbox{\Pi}^p \simeq -\frac{\pi\eta_o d^3}{4V}N (N^2-1)
[\hat{\bbox{e}}_z\hat{\bbox{e}}_z+(3\lambda N)^{-1}
(\bbox{I}-\hat{\bbox{e}}_z\hat{\bbox{e}}_z)]\cdot\bbox{\beta}.
\end{equation}

\noindent The term proportional to $(\lambda N)^{-1}$ constitutes the
first correction to the rigid chain limit ($\lambda^{-1}\rightarrow
0$). Depending upon the structure of the flow rate, this term
represents either an increase or a decrease in the value of the
different components of the pressure tensor as a function of
$\lambda$, i.e. when the chain becomes more flexible. Note that, as we
are not taking into account hydrodynamic interactions between the
particles ({\it free draining approximation}), the rigid
chain-limiting pressure tensor grows simply as the third power of the
chain length.  Hydrodynamic interactions are responsible for the
logarithmic term that should also appear when dealing with a long
straight line of spheres \cite{doi89}, but their effects are less
important for shorter chains as the ones considered here.

Moreover, from this expression we can obtain the contribution of
the chain to the viscosity tensor, $\bbox{\eta}_p$, which
we can identify by comparing with  the relationship

\begin{equation}\label{ch21}
\bbox{\Pi}^p=- \bbox{\eta}_p:\bbox{\beta}. 
\end{equation}

\noindent This tensor is thus

\begin{equation}\label{ch22}
\bbox{\eta}_p \simeq \frac{\pi\eta_o 
d^3}{4V}N (N^2-1)\left\{\hat{\bbox{e}}_z \bbox{I}
\hat{\bbox{e}}_z+(3\lambda N)^{-1} 
(\bbox{S}-\hat{\bbox{e}}_z \bbox{I} \hat{\bbox{e}}_z)\right\},
\end{equation}
 
\noindent where $\bbox{S}$ is the fourth-rank tensor $S_{ijkl}=(\delta_{ik}
\delta_{jl}+\delta_{il}\delta_{jk})/2$, symmetric in any of
these index transformations ($i\leftrightarrow j$, $k\leftrightarrow
l$, $(ij)\leftrightarrow (kl)$).

For the conditions under consideration, the chains are always oriented
in the direction of the field, but smoothly vibrate and oscillate
around this orientation. In this case, the symmetries of the fluid
flow essentially determine the characteristics of both the pressure
and the viscosity tensors. For instance, as the chains are oriented along the
z-axis---parallel to the external field---if the flow field rate
$\bbox{\beta}$ is diagonal, so will be the pressure tensor. On the other 
hand, for a non-diagonal flow rate, we have, in
general, both symmetric and antisymmetric contributions. Let us
illustrate the structure of the pressure tensor by considering some
representative cases of interest:

i) {\em Flow through a pore}. In this case, if the system has
rotational symmetry around the $z$-axis, the velocity field is
given by

\begin{equation}\label{ch23}
v_x=-\frac{1}{2}\beta x \qquad v_y=-\frac{1}{2}\beta y 
\qquad v_z=\beta z,
\end{equation}

\noindent where $\beta$ is the elongational rate. The pressure tensor
has then the following form
 
\begin{equation}\label{ch23b}
\bbox{\Pi}^p\simeq -\frac{\pi\eta_o d^3}{4V}N (N^2-1) 
\left(\begin{array}{ccc}
-(6 \lambda N)^{-1}& 0 & 0 
\\0 & -(6 \lambda N)^{-1} & 0 \\0 & 0 & 1 
\end{array}\right)\beta.
\end{equation}

\noindent Under these conditions the chain is subjected to tensile and 
compressive forces giving rise to the so-called {\it elongational
  viscosities} defined from the differences between two of the
diagonal components of the pressure tensor, i.e.  normal pressure
differences. As an example,
 
\begin{equation} \label{ch24}
-\frac{\Pi_{zz}^p-\Pi_{xx}^p}{\beta}\simeq \frac{\pi\eta_o
  d^3}{4V}N (N^2-1)[1+(6\lambda N)^{-1}].
\end{equation}

\noindent This quantity essentially grows with the third power of the 
rigid chain length $(Nd)^3$, and for high but finite values of
$\lambda$. When $\lambda^{-1}\neq 0$, the chain becomes slightly more
flexible, its effective length increases giving rise to a small
increment in the viscosity of the system.

Alternatively, a fluid flow with rotational symmetry around the $x$ (or $y$)
axis leads to the following structure of the pressure tensor:

\begin{equation}\label{ch25}
\bbox{\Pi}^p\simeq -\frac{\pi\eta_o d^3}{8V}N (N^2-1) 
\left(\begin{array}{ccc}
2(3\lambda N)^{-1} & 0 & 0 
\\0 & -(3\lambda N)^{-1} & 0 \\0 & 0 & -1 
\end{array}\right)\beta.
\end{equation}

\noindent  Combinations of any two diagonal
components of this tensor amount to different elongational viscosity
coefficients.  Moreover, in any of the above mentioned geometries, we
have another viscosity coefficient coming from the trace of the
pressure tensor.

ii) {\em Planar elongational flow}.  This flow can be generated by
four rotating cylinders. If we locate the cylinders such that
the elongational tensor is again a diagonal matrix, in view of
Eq.~(\ref{ch20}), the pressure tensor is also diagonal and, of course,
 symmetric. For example, if

\begin{equation}\label{ch29}
v_x=-\beta x \qquad v_z=\beta z,
\end{equation} 

\noindent we find a similar phenomenology as the one described 
above. On the other hand, in a reference frame rotated $45^\circ$ with
respect to the previous one, the elongational tensor is 

\begin{equation}\label{ch30}
\bbox{\beta}=\beta \left(\begin{array}{ccc}
-1 & 0 & 1 \\0 & 0 & 0 \\1 & 0 & 1 \end{array}\right),
\end{equation}
resulting in

\begin{equation}\label{ch30b}
\bbox{\Pi}^p \simeq -\frac{\pi\eta_o d^3}{4V}N (N^2-1) 
\left(\begin{array}{ccc}
-(3\lambda N)^{-1} & 0 & 1 
\\0 & 0 & 0 \\ (3\lambda N)^{-1} & 0 & 1 
\end{array}\right)\beta,
\end{equation}

\noindent which has both  symmetric and antisymmetric contributions. 
Related to these parts, we find not only {\it elongational
  viscosities} but also a {\it shear viscosity}, $\eta$, and a {\it
  rotational viscosity}, $\eta_r$. These last quantities can be
identified from the non-diagonal components of the symmetric and
antisymmetric parts, $\Pi_{xz}^{p (s)}$ and $\Pi_{xz}^{p (a)}$,
respectively. They turn out to be

\begin{equation} \label{ch33}
\eta \simeq \frac{\pi\eta_o d^3}{16V}N (N^2-1)
[1+(3\lambda N)^{-1}], 
\end{equation}
\begin{equation} \label{ch34}
\eta_r \simeq \frac{\pi\eta_o d^3}{16V}N (N^2-1)
[1-(3\lambda N)^{-1}].
\end{equation}

\begin{figure}[t]
\centerline{\epsfig{file=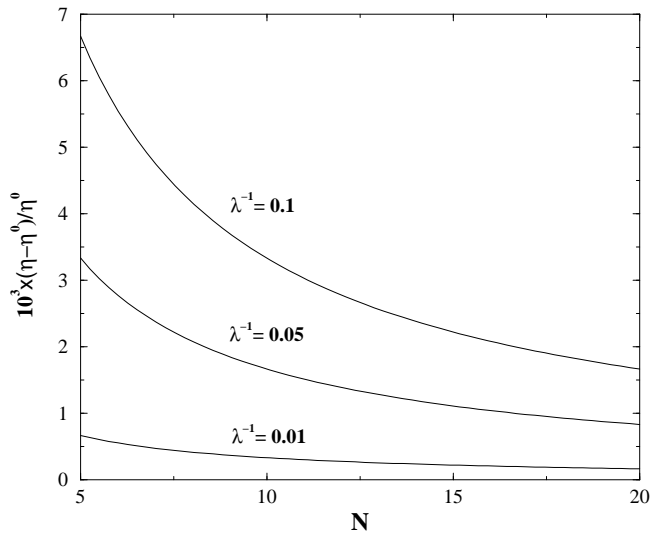, width=8.5cm}} 
\vspace*{0.5cm}
\caption{Relative correction to the shear viscosity of a chain as a
function of the number of dipolar particles $N$, for fixed values of
$\lambda^{-1}=0.01,0.05,0.1$.}
\label{fig2}
\end{figure}

\noindent Thus, for $\lambda^{-1} \rightarrow 0$, both shear and
rotational viscosities behave like those of a dilute suspension of
cylinders and grow with the third power of the rigid chain length,
$(Nd)^3$.  For high but finite values of $\lambda$, however, the shear
viscosity increases, whereas the rotational viscosity decreases. The
reason why the latter occurs is that in this particular geometry, the
elongational flow not only stretches the chain but also tends to
reorient it. This rotation is opposed by the presence of the external
field; in the case of a rigid chain $(\lambda^{-1}\rightarrow 0)$, the
magnetic field impedes more effectively the reorientation and, as a
consequence, the rotational viscosity attains its maximum value.
Nonetheless, the rotational viscosity is smaller for a finite value
of the parameter $\lambda^{-1}$, or, in other words, for a slightly
more flexible chain. In Fig.~\ref{fig2} we plot the relative variation of
the shear viscosity as a function of the number of particles in the
chain $N$, for three given values of $\lambda^{-1}=0.01,0.05,0.1$
within the experimental range. If we define the shear viscosity in the
$\lambda^{-1} \rightarrow 0$ limit as $\eta^0$, the relative
correction $(\eta-\eta^0)/\eta^0=(3\lambda N)^{-1}$ grows linearly
with $\lambda^{-1}$, and becomes less important for longer chains. The
relative decrease of the rotational viscosity
$|\eta_r-\eta^0_r|/\eta^0_r$ shows the same behavior depicted in
Fig.~\ref{fig2}.

\section{Conclusions}

\label{conclu}

In summary, we have computed the contribution of a linear chain of
dipolar particles to the Kramers' pressure tensor and the viscosities
of a dilute suspension of such clusters. The field-induced chain is
oriented along a preferred direction---parallel to the external
field---but, at the same time, is exposed to both thermal effects and
the influence of an external elongational fluid flow.

As was pointed out in previous experimental studies, these linear
clusters exhibit a certain degree of flexibility despite its rod-like
appearance. At equilibrium, thermal fluctuations are responsible for
average separations between the spheres surfaces of the order of
$(5-10)\%$ of a diameter, as well as for slight angular deviations,
which amount to an effectively larger mean square end-to-end length of
these field-oriented chains. The flexibility of a chain is obviously
limited by the strength of the dipolar interactions among its
constituent colloidal particles. The dimensionless parameter $\lambda$,
comparing dipolar and thermal energies, thus becomes an essential
ingredient in our analysis.

The rigid chain-like rheologic behavior, which corresponds to the
limit $\lambda^{-1}\rightarrow 0$, is modified considerably as the
chains become more flexible, i.e., for large but finite values of
$\lambda$.  Firstly, we have characterized the mean size of the chain
obtaining the average {\em end-to-end} vector and the mean square {\em
  end-to-end} distance as a function of the parameter $\lambda^{-1}$.
Secondly, we have computed the contribution of the chain to the
pressure tensor of the system from the rheological equation of state
proposed by Kramers.  The structure of the pressure tensor has been
explicitly illustrated for different elongational flows. Furthermore,
by comparing our result to the standard linear law relating the
pressure tensor and the elongational flow rate, we have calculated the
viscosity tensor of the system.

Some of the most common viscosity coefficients characterizing a
colloidal suspension of this kind have also been obtained. We have
discussed the role played by the parameter $\lambda$ in the behavior
of these quantities for two practical situations of interest: a flow
through a pore and a planar elongational flow. In the first one, the
symmetry of the contribution of the chains to the pressure tensor of
the suspension only gives rise to elongational viscosities, defined
from the differences between any pair of diagonal components of the
pressure tensor. As the flow field effectively stretches the chain by
a global amount proportional to $\lambda^{-1}$, the elongational
viscosity also increases by a similar factor beyond its rigid
limit value.  In the second situation described, the chain's
contribution to the pressure tensor is no longer symmetric; in this
case, we have computed the so-called shear and rotational viscosities.
The former increases with $\lambda^{-1}$ as well, whereas the latter
decreases, i.e., the reorientation of the chains is slightly favored
for large but finite values of this parameter, yielding a rotational
viscosity smaller than in the case of a rigid rod.

Our results, valid up to the orders specified in the analysis, should
provide a good description of the most frequent experimental
conditions and, consequently, could be directly contrasted with
experiments.

\acknowledgements

We are grateful to Dr. R. Pastor-Satorras for his critical reading of 
the manuscript. M. C. M. was supported by a grant from the Direcci\'o 
General de Recerca (Generalitat de Catalunya) and by the NSF Grant No.
DMR-93-03667. We also acknowledge financial support by the DGICYT of
the Spanish Government (Grant No. PB95-0881). 

\end{multicols}

\appendix 

\section*{}

In this appendix, we indicate in some detail some of the steps
involved in the evaluation of Eq.~(\ref{ch14}). First, we split up the
integral into different parts, a)-e), which are

\begin{eqnarray}\label{chab1}
\mbox{a)} &\quad& \int (\prod_{l=1}^{N-1} d \bbox{q}_l\ {\rm exp}
(-\Phi_{l,l+1})) 
\left(\bbox{q}_k \frac{\partial V_{mag}}{\partial \bbox{q}_k} \right)
\\ 
\mbox{b)} &\quad& \frac{\bbox{\beta}}{2D}:\sum_{i(i\neq
  k)} \sum_{j(j\neq i,k)}  
C_{ij}\int (\prod_{l=1}^{N-1} d\bbox{q}_l\ {\rm exp} (-\Phi_{l,l+1}))\
\bbox{q}_i\  \bbox{q}_j  
\left(\bbox{q}_k \frac{\partial V_{mag}}{\partial \bbox{q}_k} \right)
\\
\mbox{c)} &\quad& \frac{\bbox{
  \beta}}{2D}:\sum_{i(i\neq k)} C_{ii}\int (\prod_{l=1}^{N-1}
  d\bbox{q}_l\ {\rm exp} (-\Phi_{l,l+1}))\ \bbox{q}_i\ \bbox{q}_i\ 
  \left(\bbox{q}_k
    \frac{\partial V_{mag}}{\partial \bbox{q}_k} \right) \\
\mbox{d)} &\quad& \frac{\bbox{\beta}}{2D}:\sum_{i(i\neq k)}
C_{ik}\int  (\prod_{l=1}^{N-1} 
d\bbox{q}_l\ {\rm exp} (-\Phi_{l,l+1}))\ \bbox{q}_i\ \bbox{q}_k\
\left(\bbox{q}_k  
\frac{\partial V_{mag}}{\partial \bbox{q}_k} \right) \\
\mbox{e)} &\quad& \frac{\bbox{\beta}}{2D}:C_{kk} \int
(\prod_{l=1}^{N-1} d\bbox{q}_l\ {\rm exp} 
(-\Phi_{l,l+1}))\ \bbox{q}_k\ \bbox{q}_k\ \left(\bbox{q}_k  
\frac{\partial V_{mag}}{\partial \bbox{q}_k} \right). 
\end{eqnarray}

Let us explicitly compute some of these contributions. For instance, 

\begin{eqnarray}
\mbox{a)} &\quad& \int (\prod_{l=1}^{N-1} d \bbox{q}_l\ {\rm
  exp}(-\Phi_{l,l+1})) \left(\bbox{q}_k  
\frac{\partial V_{mag}}{\partial \bbox{q}_k} \right) \nonumber \\
&=&\left[\int d \bbox{q}_k {\rm exp}(-\Phi_{k,k+1})\right]^{N-2} 
\int d \bbox{q}_k\ \bbox{q}_k 
\frac{\partial \Phi_{k,k+1}}{\partial \bbox{q}_k} {\rm
  exp}(-\Phi_{k,k+1})  \nonumber \\
&\simeq& \left[2\pi \int_0^{\infty} q_k^2 dq_k \int_0^{\pi}\sin
  \theta_k d \theta_k e^{\lambda(2-3\theta_k^2-6\xi_k)}\right]^{N-1} 
k_BT \bbox{I} \nonumber \\
&\simeq& \left[2\pi d^3 e^{2\lambda} \int_0^{\infty} (1+2\xi_k) d\xi_k 
e^{-6\lambda\xi_k} \int_0^{\infty} 
\left(\theta_k - \frac{\theta_k^3}{3!}\right) d\theta_k 
e^{-3\lambda\theta_k^2}\right]^{N-1} k_BT \bbox{I} \nonumber \\
&\simeq& \left[\frac{\pi d^3 e^{2\lambda}}{18\lambda^2} 
\left(1+\frac{5}{18\lambda}\right)\right]^{N-1}k_BT \bbox{I},
\end{eqnarray}

\begin{eqnarray}
\mbox{b)} &\quad& \frac{\bbox{\beta}}{2D}:\sum_{i(i\neq
  k)} \sum_{j(j\neq i,k)}  
C_{ij}\int (\prod_{l=1}^{N-1} d\bbox{q}_l\ {\rm exp}
  (-\Phi_{l,l+1}))\bbox{q}_i\  \bbox{q}_j  
\left(\bbox{q}_k \frac{\partial V_{mag}}{\partial \bbox{q}_k}
  \right) \nonumber \\  
&\simeq&\left[\frac{\pi d^3 e^{2\lambda}}{18\lambda^2}
\left(1+\frac{5}{18\lambda}\right)\right]^{N-3}\frac{\bbox{
  \beta}}{2D}: \sum_{i(i\neq k)} \sum_{j(j\neq i,k)} C_{ij} \times 
  \nonumber \\
& &\left[\int d \bbox{q}_i {\rm exp} (-\Phi_{i,i+1})\
  \bbox{q}_i\right]^2 \int d \bbox{q}_k\ \bbox{q}_k \frac{\partial
  \Phi_{k,k+1}}{\partial \bbox{q}_k} {\rm exp}(-\Phi_{k,k+1}) 
\nonumber  \\ 
&\simeq&\left[\frac{\pi d^3 e^{2\lambda}}{18\lambda^2}
\left(1+\frac{5}{18\lambda}\right)\right]^{N-1}
\frac{k_BTd^2\beta_{zz}}{2D} \sum_{i(i\neq k)} \sum_{j(j\neq i,k)}
  C_{ij} \bbox{I}.
\end{eqnarray}

\noindent  The other contributions are obtained in a similar fashion. 
Adding up the different parts and considering the relations
(\ref{ch17b}), we can rewrite the average in Eq.~(\ref{ch14}) as

\begin{eqnarray}\label{chab7}
& &\left\langle \bbox{q}_k 
\frac{\partial V_{mag}}{\partial \bbox{q}_k} \right\rangle= C
\left[\frac{\pi d^3 e^{2\lambda}}{18\lambda^2} 
\left(1+\frac{5}{18\lambda}\right)\right]^{N-1} k_BT \left\{\left[1+
\frac{d^2 \beta_{zz}}{24D}N(N^2-1)\right]\bbox{I} \right.\nonumber \\
& & \mbox{\ \ } \left. + \frac{d^2}{D}\sum_{i} C_{ik}\ 
\bbox{\beta}\cdot\hat{\bbox{e}}_z\hat{\bbox{e}}_z+
\frac{d^2}{12D}\lambda^{-1} 
\left[\bbox{\beta}:(\bbox{I}-\hat{\bbox{e}}_z\hat{\bbox{e}}_z)\frac{N^2-1}{6}
\bbox{I} + 2 C_{kk}\bbox{\beta}\cdot(\bbox{I}-
\hat{\bbox{e}}_z\hat{\bbox{e}}_z) \right] \right\}.
\end{eqnarray}

Finally, after the appropriate normalization of the probability
density, and up to first order in $\bbox{\beta}$, we obtain

\begin{eqnarray}\label{chab8}
\left\langle \bbox{q}_k 
\frac{\partial V_{mag}}{\partial \bbox{q}_k} \right\rangle&=& k_BT
\left\{\left[1+ \frac{d^2 \beta_{zz}}{24D}N(N^2 -1)\right]
\bbox{I}+ \frac{d^2}{D}\sum_{i} C_{ik}\ \bbox{\beta}
\cdot\hat{\bbox{e}}_z\hat{\bbox{e}}_z \right. \nonumber \\
&+& \left. \frac{d^2}{12D}\lambda^{-1} 
\left[\bbox{\beta}:(\bbox{I}-\hat{\bbox{e}}_z\hat{\bbox{e}}_z)
 \frac{N^2-1}{6} \bbox{I} + 2 C_{kk}\ \bbox{
\beta}\cdot(\bbox{I}-\hat{\bbox{e}}_z\hat{\bbox{e}}_z)\right] \right\}
\nonumber \\
&\times& \left\{1-\frac{d^2 \beta_{zz}}{24D} N(N^2 -1) -
  \frac{d^2}{72D}\lambda^{-1} (N^2 -1)\ \bbox{
\beta}: (\bbox{I}-\hat{\bbox{e}}_z\hat{\bbox{e}}_z)\right\},
\end{eqnarray}

\noindent which can be simplified to the final form (\ref{ch19}).

\begin{multicols}{2}

\end{multicols}


\begin{references} 
  
\bibitem{gast89} 
A.P. Gast and C.F. Zukoski, Adv. Colloid Interface Sci. {\bf 30}, 
153 (1989).

\bibitem{VdV89} 
T.G.M. van de Ven, {\it Colloidal Hydrodynamics},
  (Academic Press, London, 1989).
  
\bibitem{shliomis72} M.I. Shliomis, Sov. Phys. JETP {\bf 34}, 1291
  (1972).
  
\bibitem{miguel95} 
M.C. Miguel and M. Rub\'{\i}, Phys. Rev. E {\bf 51}, 2190 (1995).
  
\bibitem{bacri95} 
J.-C. Bacri, R. Perzynski, M.I. Shliomis and G.I. Burde, 
Phys. Rev. Lett.  {\bf 75}, 2128 (1995).

\bibitem{winslow49}
W.M. Winslow, J. Appl. Phys. {\bf 20}, 1137 (1949).

\bibitem{bossis90}
G. Bossis, C. Mathis, Z. Mimouni, and C. Paparoditis,
Europhys. Lett. {\bf 11}, 133 (1990). 

\bibitem{skjeltorp85}
A.T. Skjeltorp, Phys. Rev. Lett. {\bf 51}, 2306 (1983); J. Magn. 
Magn. Mat.  {\bf 65}, 195 (1987); J. Appl. Phys. {\bf 57} (1), 
 3285 (1985).
 
\bibitem{davies86}
 P. Davies, J. Popplewell, G. Martin, A. Bradbury and R.W. 
Chantrell, J.  Phys. D, Appl. Phys. {\bf 19}, 469 (1986). 

\bibitem{ferrofluid} In a ferrofluid one observes similar
  phenomena but, due to fact that the particles are permanently
  magnetized, the aggregation takes place even without an applied
  magnetic field.

\bibitem{miyazima87}
S. Miyazima, P. Meakin, and F. Family, Phys. Rev. A {\bf 36}, 1421
(1987). 


\bibitem{helgesen88}
G. Helgesen, A.T. Skjeltorp, P.M. Mors, R. Botet, and R. Jullien,
Phys. Rev. Lett. {\bf 61}, 1736 (1988).


\bibitem{bossis89}
G. Bossis and J.F. Brady, J. Chem. Phys. {\bf 91}, 1866 (1989).

\bibitem{fraden89}
S. Fraden, A.J. Hurd, and R.B. Meyer, Phys. Rev. Lett. {\bf 63}, 2373
(1989).

\bibitem{fermigier92} M. Fermigier and A.P. Gast, J. Colloid and
  Interface Sci. {\bf 154}, 522 (1992).  
  
\bibitem{klingenberg93}
D.J. Klingenberg, C.F. Zukoski, and J.C. Hill, J. Appl. Phys. {\bf
  73}, 4644 (1993).

\bibitem{lemaire94} E. Lemaire, Y. Grasselli and G. Bossis, J. Phys.
  II France {\bf 2}, 359 (1992); J. Phys. II France {\bf 4}, 253
  (1994).

\bibitem{gast95}
J.H.E. Promislow, A.P. Gast, and M. Fermigier, J. Chem. Phys. {\bf
  102}, 5492 (1995).

\bibitem{miguel98} M.C. Miguel and R. Pastor-Satorras, submitted to
  Phys. Rev. E, (1998).

\bibitem{deGennes70}
P.G. de Gennes and P. Pincus, Phys. Kondens. Mater. {\bf 11}, 189
(1970). 

\bibitem{wang94}
H. Wang, Y. Zhu, C. Boyd, W. Luo, A. Cebers, and R.E. Rosensweig,
Phys. Rev. Lett. {\bf 72}, 1929 (1994). 

\bibitem{jund95}
P. Jund, S.G. Kim, D. Tom\'{a}nek and J. Hetherington,
Phys. Rev. Lett. {\bf 74}, 3049 (1995).
 
\bibitem{ferrofluid2} In a suspension of superparamagnetic particles,
  the induced moment is given by $m=\chi H V_p$, with $\chi$ the
  effective magnetic susceptibility of the particle. For the
  monodomain magnetic particles constituting a ferrofluid $m=M_s V_p$,
  with $M_s$ the saturation magnetization of the ferromagnetic
  material, and $V_p$ the particle volume.  In the latter case, 
  only when $\mu\rightarrow \infty$ all dipoles will be aligned 
  parallel to the external field.

\bibitem{rotation} We disregard rotational contributions to the
diffusion equation. In the presence of a strong external field, the
magnetic moment, rigidly attached to a ferromagnetic particle, is
oriented along the field direction, hindering the particle
rotation. This is no longer the case in a magnetorheological fluid
containing field-induced dipolar particles. Nevertheless, in the ideal
situation of an induced moment completely detached from the body of
the host particle, and in the absence of hydrodynamic interactions,
the translational and rotational dynamics of such spheres are
decoupled. 
  
\bibitem{bird76}
 R.B. Bird, O. Hassager, R.C. Armstrong and C.F.Curtis, 
 {\it Dynamics of Polymeric Liquids}, (Wiley, New York,
  1976).

\bibitem{rigid} As for polymers, one can introduce a correlation or
  persistence length $\lambda_p\sim \lambda d$. A rigid chain will
  then be characterized by $\lambda_p\gg Nd$, or $\lambda \gg N$.

\bibitem{doi89}
M. Doi and S.F. Edwards, {\it The Theory of Polymer Dynamics}, 
(Oxford University Press, New York, 1989).

\end{references}
\end{document}